\documentclass[twocolumn,aps,showpacs,preprintnumbers,amsmath,amssymb, superscriptaddress]{revtex4-1}
\pdfoutput=1

\usepackage{bm}
\usepackage{graphicx}
\usepackage{color}

\renewcommand{\vec}{\mathbf}

\begin{document}

\title{Raman study of lattice dynamics in Weyl semimetal TaAs}

\author{H. W. Liu}\email{hliu@iphy.ac.cn}
\affiliation{Beijing National Laboratory for Condensed Matter Physics, and Institute of Physics, Chinese Academy of Sciences, Beijing 100190, China}
\author{P. Richard}\email{p.richard@iphy.ac.cn}
\affiliation{Beijing National Laboratory for Condensed Matter Physics, and Institute of Physics, Chinese Academy of Sciences, Beijing 100190, China}
\affiliation{Collaborative Innovation Center of Quantum Matter, Beijing, China}
\author{Z. D. Song}
\affiliation{Beijing National Laboratory for Condensed Matter Physics, and Institute of Physics, Chinese Academy of Sciences, Beijing 100190, China}
\author{L. X. Zhao}
\affiliation{Beijing National Laboratory for Condensed Matter Physics, and Institute of Physics, Chinese Academy of Sciences, Beijing 100190, China}
\author{Z. Fang}
\affiliation{Beijing National Laboratory for Condensed Matter Physics, and Institute of Physics, Chinese Academy of Sciences, Beijing 100190, China}
\affiliation{Collaborative Innovation Center of Quantum Matter, Beijing, China}
\author{G.-F. Chen}
\affiliation{Beijing National Laboratory for Condensed Matter Physics, and Institute of Physics, Chinese Academy of Sciences, Beijing 100190, China}
\affiliation{Collaborative Innovation Center of Quantum Matter, Beijing, China}
\author{H. Ding}
\affiliation{Beijing National Laboratory for Condensed Matter Physics, and Institute of Physics, Chinese Academy of Sciences, Beijing 100190, China}
\affiliation{Collaborative Innovation Center of Quantum Matter, Beijing, China}

\date{\today}

%\begin{minipage}[t]{6.8in}
\begin{abstract}
We report a polarized Raman study of Weyl semimetal TaAs. We observe all the optical phonons, with energies and symmetries consistent with our first-principles calculations. We detect additional excitations assigned to multiple-phonon excitations. These excitations are accompanied by broad peaks separated by 140~cm$^{-1}$ that are also most likely associated with multiple-phonon excitations. We also noticed a sizable B$_1$ component for the spectral background, for which the origin remains unclear.
\end{abstract}


%74.70.Xa 	Pnictides and chalcogenides 
%74.25.Jb 	Electronic structure (photoemission, etc.) 
%79.60.-i 	Photoemission and photoelectron spectra
%78.30.-j 	Infrared and Raman spectra 
%63.20.-e 	Phonons in crystal lattices

%\end{minipage}
\maketitle
%\narrowtext

\section{Introduction}

Although predicted long ago as a solution of the Weyl equation describing relativistic spin 1/2 fermions \cite{Weyl1929}, Weyl fermions have never been observed experimentally in high-energy physics. A few years ago the existence of Weyl nodes behaving like Weyl fermions has been predicted in condensed matter physics \cite{Balents_Physics2011,G_Xu_PRL107}, in materials called Weyl semimetals. An isolated Weyl node can be assimilated in momentum space to a monopole of topological charge defined by its chirality \cite{G_Xu_PRL107}. However, only very recently did the existence of Weyl nodes proven experimentally. In particular, the existence of Weyl nodes predicted in TaAs \cite{H_Weng_PRX5} has been inferred experimentally from its topological surface state measured by ARPES \cite{Lv_BQ_arc} and from its negative magneto-resistance induced by the chiral anomaly \cite{X_Huang_arxiv}, and more directly from its bulk electronic band structure determined from bulk-sensitive ARPES measurements \cite{Lv_BQ_bulk}. Besides the interesting physics derived from its topological surface and the topological nature of Weyl nodes, TaAs has a great potential for applications. Indeed, not only it is a binary compound with a relatively simple crystal structure, the Weyl nodes are associated with bulk properties, in contrast to topological insulators for which only the surface shows a particular interest. Despite a complete description of the electronic structure of this exotic material, there is only little known, both theoretically and experimentally, about the lattice dynamics of TaAs. 

In this paper, we report a Raman study of TaAs. We observe all the optical modes of TaAs at the Brillouin zone (BZ) center ($\Gamma$), which consist in one A$_1$ mode, two B$_1$ modes and three E modes. The experimental energies of these modes, along with their symmetries, are consistent with our first-principles calculations. We detect peaks beyond the range of the single-phonon excitations assigned to multiple-phonon excitations, as well as broader features separated by 140~cm$^{-1}$ that also likely correspond to multiple-phonon excitations. Finally, we report a sizable B$_1$ component for the spectral background, which has an undetermined origin.

\section{Experiment}

The single crystals of TaAs used in this study were grown by chemical vapor transport and characterized by X-ray diffraction to determine the crystal orientation \cite{X_Huang_arxiv}. Freshly prepared platelike samples with typical size of $0.4\times 0.4\times 0.08$ mm$^3$ were prepared for Raman scattering measurements. The measurements were performed with the 514.5~nm and 488.0~nm excitations of an Ar-Kr laser focussed on flat sample surface regions with a 100$\times$ objective mounted in a back-scattering micro-Raman configuration. The power at the sample was smaller than 0.4~mW. The signal was analyzed by a Horiba Jobin Yvon T64000 spectrometer equipped with a nitrogen-cooled CCD camera. 

\section{Calculations}

The non-centrosymmetric structure of TaAs is characterized by the space group I41md ($C_{4v}^{11}$, group no. 109), with 4 atoms in one unit cell. An analysis in terms of the irreducible representations of this group shows that the vibration modes of this system decompose \cite{Comarou_Bilbao} into [A$_1$+E]+[A$_1$+2B$_1$+3E], where the first and second terms correspond to the acoustic and optic phonon modes, respectively. While all the optic modes are Raman (R) active, only the A$_1$ and E modes are infrared (IR) active. To determine the phonon vibrational configurations and estimate the phonon frequencies, we performed calculations using the first-principles pseudopotential plane wave method package Quantum Espresso \cite{Giannozzi_JPCM21}. We set a $12\times 12\times 12$ monkhorst-pack momentum ($k$) point mesh and a 45~Ry cutoff for the electronic wavefunctions. The exchange and correlation functional was treated within the generalized gradient approximation (GGA) of Perdew-Burke-Ernzerhof \cite{Perdew_PRL77}. Coordinates and cell shape of experimental data \cite{Furuseth_ACS19} have been fully relaxed until the forces acting on atoms are all smaller than 10$^{-4}$~Ry/aB and the pressure is smaller than 0.2~kbar. Using the information on the ground state, it is easy to use the Phonon package, which implements the density functional perturbation theory (DFPT) \cite{Baroni_PRL58,Baroni_RMP73,Gonze_PRA52} to get the phonon frequencies and vibration modes at the $\Gamma$ point. All the optic vibration modes of TaAs are illustrated in Fig.~\ref{modes}. For each mode we indicate the corresponding irreducible representation, the experimental and calculated phonon frequencies, as well as the optical activity. While the E modes are related to vibrations in the ab plane, the A$_1$ and B$_1$ modes are associated with vibrations along the c axis. In addition, we also calculated the phonon dispersion along high-symmetry lines in the momentum space. We first applied DFPT calculations using the Phonon package on a coarse $4\times 4\times 4$ monkharst-pack phonon momentum $(q)$ point mesh to get their frequencies and modes. The dynamic matrix in real space is then obtained by Fourier-transformation of these modes. Finally, we transformed this real space dynamic matrix back to general $q$ points and diagonalized it to get the dispersion. 

\section{Results and discussion}

Experimentally, the symmetry of the vibration modes can be described by the relevant Raman tensors of the C$_{4v}$ group (note that the A$_2$ channel is not Raman active):
\begin{displaymath}
\textrm{A$_1$}=
\left(\begin{array}{ccc}
a & 0 &0\\
0 & a &0\\
0 & 0 &b
\end{array}\right), 
\textrm{B$_{1}$ =}
\left(\begin{array}{ccc}
c & 0 &0\\
0 & -c &0\\
0 & 0 &0\\
\end{array}\right),
\textrm{B$_{2}$ =}
\left(\begin{array}{ccc}
0 & d &0\\
d & 0 &0\\
0 & 0 &0\\
\end{array}\right),
\end{displaymath}

\begin{displaymath}
\left\{
\textrm{E$(x)$}=
\left(\begin{array}{ccc}
0 & 0 &e\\
0 & 0 &0\\
e & 0 &0
\end{array}\right)
, \textrm{E$(y)$}=
\left(\begin{array}{ccc}
0 & 0 &0\\
0 & 0 &e\\
0 & e &0\\
\end{array}\right)\right\}.
\end{displaymath}
\noindent 

\begin{figure}[!t]
\begin{center}
\includegraphics[width=3.4in]{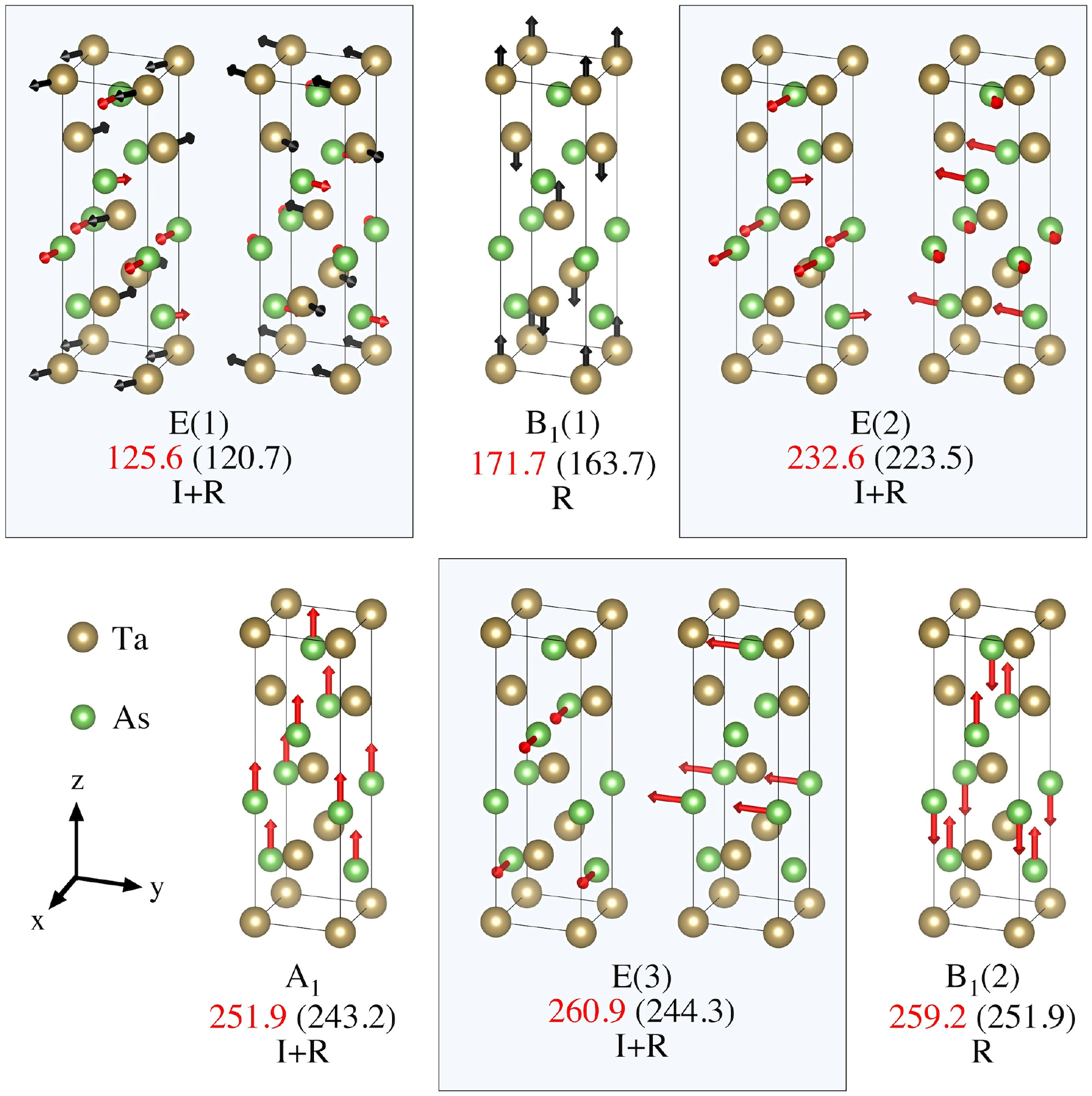}
\end{center}
\caption{\label{modes}(Color online). Main atomic displacements for the optical modes of TaAs. The displacements of Ta and As atoms are indicated by black and red arrows, respectively. We caution that the displacements of Ta atoms are not represented when they are found too small compared to the displacements of As atoms. The first line below each configuration of vibration indicates the mode symmetry in the C$_{4v}$ group notation. Numbers in parenthesis are used to specify modes when there are more than one mode with the same symmetry. The second line below each configuration of vibration indicates the experimental(calculated) mode energy. The last line below each configuration of vibration gives the optical activity, with I = infrared active, R~=~Raman active and I+R = infrared and Raman active.}
\end{figure}

\begin{figure*}[!t]
\begin{center}
\includegraphics[width=18cm]{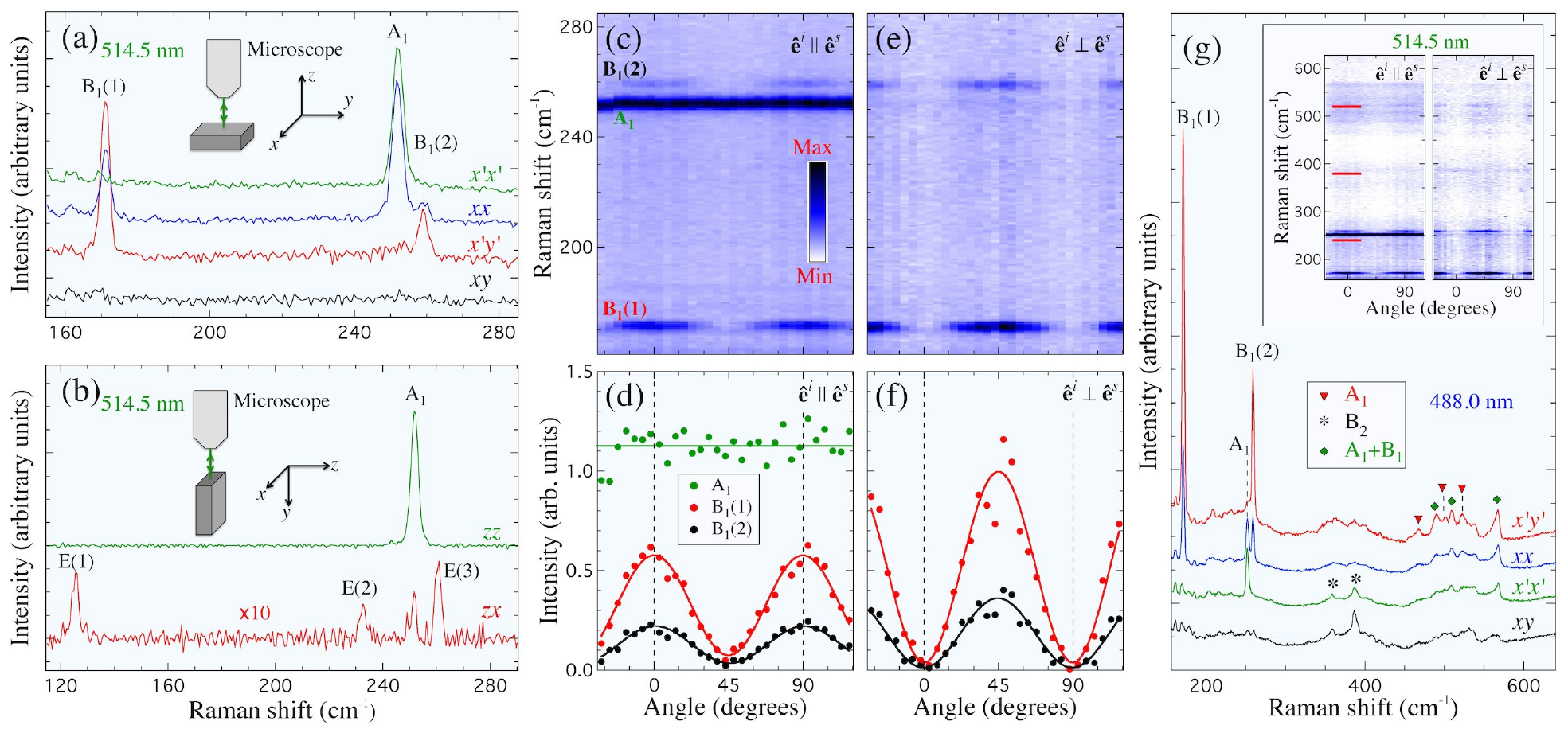}
\end{center}
\caption{\label{RT}(Color online) (a) Room temperature Raman spectra of TaAs for various ab-plane configurations of the incident and scattered light polarizations. The inset illustrates the experimental configuration. (b) Same as (a) but for the ac plane. The background of the ac spectra has been removed and the spectrum in the $zx$ configuration has been multiplied by 10 for a better visualization. (c) Intensity plot of the Raman shift as a function of the in-plane angle defined with respect to the $x$ axis, for $\vec{\hat{e}}^{i}||\vec{\hat{e}}^{s}$. (d) Angular dependence of the intensity of the Raman modes observed in (c) with $\vec{\hat{e}}^{i}||\vec{\hat{e}}^{s}$. (e) Same as (c) but for $\vec{\hat{e}}^{i}\perp\vec{\hat{e}}^{s}$. (f) Same as (d) but for the Raman modes observed in (e) with $\vec{\hat{e}}^{i}\perp\vec{\hat{e}}^{s}$. (g) Long-range spectra of TaAs recorded with 488.0~nm laser excitation in several ab-plane configurations of polarization. The dominant symmetries of some sharp excitations at high energy are indicated. The inset shows the ab-plane angular dependence of the Raman intensity recorded with 514.5~nm laser excitation under the $\vec{\hat{e}}^{i}||\vec{\hat{e}}^{s}$ and $\vec{\hat{e}}^{i}\perp\vec{\hat{e}}^{s}$ configurations of polarization. The red horizontal lines indicate the broad humps at 240~cm$^{-1}$, 380~cm$^{-1}$ and 520~cm$^{-1}$. The contrast has been adapted to illustrate a sizable B$_1$ symmetry component for the spectral background.} 
\end{figure*}

In Fig.~\ref{RT} we show the Raman spectra obtained in different ab-plane configurations of incident ($\vec{\hat{e}}^{i}$) and scattered ($\vec{\hat{e}}^{s}$) polarizations. Pure A$_1$ and B$_1$ symmetries are obtained in the $x'x'$ and $x'y'$ configurations of polarizations, respectively. As expected from our group theory analysis, only one mode is observed in the $x'x'$ spectrum. Its energy, 251.9 cm$^{-1}$, is slightly higher than the calculated value (243.2 cm$^{-1}$). Similarly, we detected two peaks in the $x'y'$ configuration, in agreement with our analysis. The B$_1(1)$ peak is observed at 171.7 cm$^{-1}$ while the B$_1(2)$ peak is detected at 259.2 cm$^{-1}$. In agreement with their symmetries, both the A$_1$ and the two B$_1$ modes are observed in the $xx$ configuration while none of them appears in the $xy$ spectrum.

In order to confirm further the symmetry of the Raman peaks observed, we performed an angular dependence study of the ab spectra. In Figs.~\ref{RT}(c) and \ref{RT}(e) we display the intensity plots of the Raman spectra as a function of the in-plane angle $\theta$ defined with respect to the $x$ axis for $\vec{\hat{e}}^{i}||\vec{\hat{e}}^{s}$ and $\vec{\hat{e}}^{i}\perp\vec{\hat{e}}^{s}$, respectively. The intensity of the A$_1$ and B$_1$ modes for the corresponding plots are given in Figs.~\ref{RT}(d) and \ref{RT}(f), respectively. As expected, the A$_1$ peak is observed only clearly for $\vec{\hat{e}}^{i}||\vec{\hat{e}}^{s}$, with an intensity insensitive to the angle, as shown in Fig.~\ref{RT}(f). In contrast, the B$_1$ modes exhibit a strong angular dependence for both the $\vec{\hat{e}}^{i}||\vec{\hat{e}}^{s}$ and $\vec{\hat{e}}^{i}\perp\vec{\hat{e}}^{s}$ configurations. For $\vec{\hat{e}}^{i}||\vec{\hat{e}}^{s}$, we expect from the B$_1$ Raman tensor that the intensity should vary with angle like $|c\cos (2\theta)|^2$, with a four-fold symmetry characterized by a maximum at 0 degrees and a node at 45 degrees, which is consistent with our observation, as shown in Fig.~\ref{RT}(d). For $\vec{\hat{e}}^{i}\perp\vec{\hat{e}}^{s}$, this pattern is shifted by 45 degrees as the B$_1$ Raman tensor predicts an angular variation of the intensity proportional to $|c\sin (2\theta)|^2$, in agreement with our observation (see Fig.~\ref{RT}(f)). Although this cannot be seen clearly from the individual spectra, we also point out a very small leak of the A$_1$ mode into the image plot displayed in Fig.~\ref{RT}(e), obtained with $\vec{\hat{e}}^{i}\perp\vec{\hat{e}}^{s}$. We attribute this leak to a small misalignment of the relative angle between $\vec{\hat{e}}^{i}$ and $\vec{\hat{e}}^{s}$ away from perfect right angle.

We now describe Raman measurements of the ac plane. For perfect alignment, no B$_1$ phonon should be observed while only the A$_1$ phonon should be detected when $\vec{\hat{e}}^{i}||\vec{\hat{e}}^{s}$. This is consistent with the $zz$ spectrum displayed in Fig.~\ref{RT}(b), which exhibits the A$_1$ mode at the same energy as in the ab plane spectra. Despite a much smaller intensity, we can detect some peaks in the $zx$ configuration. In principle, only E modes should be observed in this configuration. However, four peaks (confirmed in several measurements) are detected instead of the three predicted. By matching the experimental data with the calculated values, we can unambiguously identify the E(1) and E(2) modes at 125.6 cm$^{-1}$ and 232.6 cm$^{-1}$, respectively. This leaves two candidate peaks at 251.7 cm$^{-1}$ and 260.9 cm$^{-1}$ to be assigned to the E(3) phonon, calculated to be located at 244.3 cm$^{-1}$, as shown in Fig.~\ref{modes}. The most likely explanation for the origin of the 251.7 cm$^{-1}$ is its assignment to the A$_1$ mode, which has the same energy within error bars. Although this would be strictly forbidden for cross polarizations, we explained above a possible misalignment of the $\vec{\hat{e}}^{i}$ and $\vec{\hat{e}}^{s}$ polarizations away from perfect right angle. Since the A$_1$ is much stronger in the $zz$ spectrum as compared to the Raman intensity recorded in the $zx$ spectrum, this hypothesis is quite plausible. Following this conclusion, we tentatively assign the peak at 260.9 cm$^{-1}$ to the E(3) phonon. Although this energy is close to that of the B$_1$(2) mode (259.2 cm$^{-1}$), the energy difference is measurable. 

\begin{figure}[!t]
\begin{center}
\includegraphics[width=3.4in]{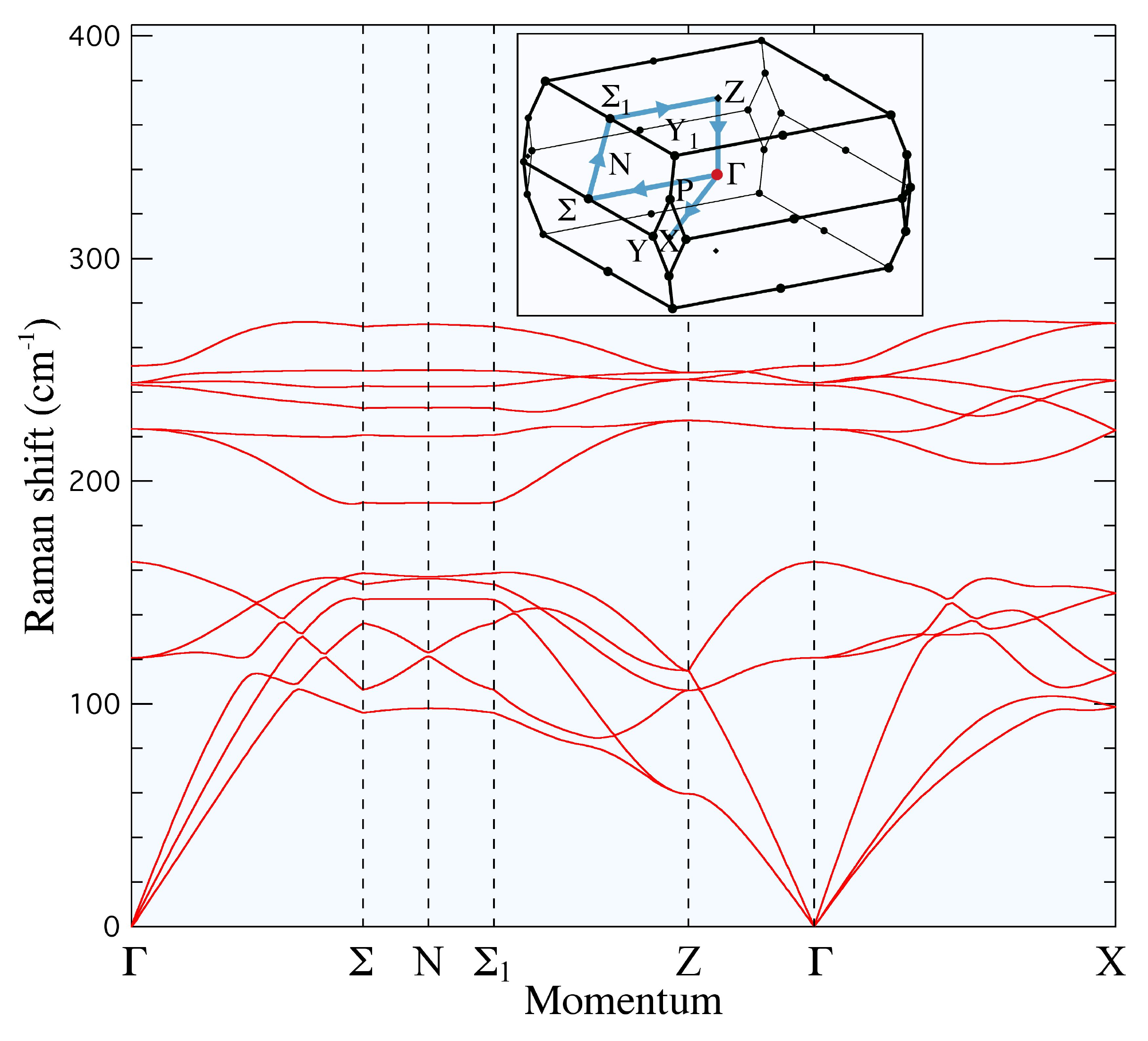}
\end{center}
\caption{\label{k_phonons}(Color online). Calculation of the phonon band dispersions in TaAs along high-symmetry lines. The inset gives the definition of high-symmetry points in the momentum space. The blue arrows show the momentum path corresponding to the dispersion displayed.}
\end{figure}

So far we discussed single-phonon excitations, which have decent agreement with our calculations. The upper energy limit is 260.9~cm$^{-1}$. However, many Raman excitations are observed beyond that limit, as shown in Fig.~\ref{RT}(g). Considering that the phonon modes disperse in the momentum space, and allowing for small discrepancies between experiments and calculations, the energy range where these extra excitations are detected is consistent with double-phonon excitations, as confirmed by our calculation of the phonon dispersion along high-symmetry lines displayed in Fig.~\ref{k_phonons}. It is worth noticing at this point that these additional excitations are observed for all the samples that we measured, and also with the two laser wavelengths excitations used in our study. For example, Fig.~\ref{RT}(g) clearly shows their existence under 488.0~nm excitation while its inset shows clearly some intensity in the same energy range, as measured under 514.5~nm excitation. By comparing the spectra in Figs.~\ref{RT}(a) and \ref{RT}(g), we report a variation of the relative intensity of the A$_1$ and B$_1$ phonons as the laser excitation varies. In contrast to the measurements performed using a 514.5~nm laser wavelength, for which the spectra show a stronger intensity of the A$_1$ peaks as compared to the B$_1$ peaks, the intensity of the B$_1$ peaks becomes larger than that of the A$_1$ mode under 488.0~nm laser excitation.

Double-phonon excitations can occur not only at the $\Gamma$ point, but also anywhere on the BZ boundary, with the only restriction that the total momentum is 0. A complete assignment of each feature is rather complicated and beyond the scope of our study, not only because modes disperse but also because their symmetries vary with momentum. Nevertheless, the symmetries of many excitations can be described in terms of the C$_{4v}$ point group and at least two of them can be tentatively identified as double excitations involving phonons at $\Gamma$. Indeed, we observe two sharp peaks at 358.8 cm$^{-1}$ and 386.7 cm$^{-1}$ that appear clearly in the $x'x'$ and $xy$ spectra. Such behavior is expected for a B$_2$ excitation. We also notice some reminiscence of that peak in the $x'y'$ spectrum, corresponding to a A$_1$ symmetry. As the ground state coincides with the A$_1$ representation, the symmetry of the excitation is directly determined by the tensor product of the irreproducible representations characterizing the two phonon modes involved. The only three possibilities to get a double-phonon excitation at $\Gamma$ with a B$_2$ symmetry are: 1) A$_1$~$\times$~B$_2$~=~B$_2$; 2) A$_2$~$\times$~B$_1$~=~B$_2$; 3) E~$\times$~E~=~A$_1$~+~A$_2$~+~B$_1$~+~B$_2$. Since there is no A$_2$ and B$_2$ mode at $\Gamma$, the only possibility is to combine E modes. Interestingly, the excitations at 358.8~cm$^{-1}$ and 386.7~cm$^{-1}$ coincide almost perfectly with [E(1)~+~E(2)] (125.6~cm$^{-1}$~+~232.6~cm$^{-1}$~=~358.2~cm$^{-1}$) and [E(1)~+~E(3)] (125.6~cm$^{-1}$~+~260.9~cm$^{-1}$~=~386.5~cm$^{-1}$), respectively. 

We also observe additional features most likely related to double-phonon excitations, although a precise assignment of the modes involved and the corresponding momentum locations remains uncertain. Among peaks with a dominant but not necessarily exclusive A$_1$ symmetry (as described within the C$_{4v}$ point group), we detect features at 468.5~cm$^{-1}$, 501.7~cm$^{-1}$ and 523.1~cm$^{-1}$. Similarly, the peaks at 490.2~cm$^{-1}$, 509.2~cm$^{-1}$ and 567.7~cm$^{-1}$ have a dominant A$_1$~+~B$_1$ symmetry. Finally, we notice some broader features with a full width of about 100~cm$^{-1}$. The centers of these broad features are located at 240~cm$^{-1}$, 380~cm$^{-1}$ and 520~cm$^{-1}$, as shown in Fig.~\ref{RT}(g) and its inset. Interestingly, these broad humps are separated by 140~cm$^{-1}$. These features are detected in all the ab-plane configurations of polarization. 

Interestingly, the inset of Fig.~\ref{RT}(g) suggests that the spectral background itself is carries a sizable B$_1$ symmetry component. Indeed, we observe an intensity modulation with angle that is similar to that of a B$_1$ phonon, with lower intensity at 45 degrees for $\vec{\hat{e}}^{i}||\vec{\hat{e}}^{s}$, and at 0 and 90 degrees with $\vec{\hat{e}}^{i}\perp\vec{\hat{e}}^{s}$. Whether this phenomenon origins from lattice vibration or is related the Weyl semimetal nature of the electronic structure of TaAs, is unclear.

\section{Summary}

In summary, we performed a polarized Raman study of Weyl semimetal TaAs. We identified all the optic phonon modes of this material. Their energies and symmetries are consistent with our first-principles calculations. We observed additional features assigned to double-phonon excitations, as well as much larger humps separated by 140~cm$^{-1}$ that are also likely to have a multiple-phonon origin. Finally, we identified a sizable B$_1$ component for the spectral background, which origin is unknown.

\section*{Acknowledgement}

We acknowledge W.-L. Zhang, S. F. Wu, D. Chen and H. M. Weng for useful discussions. This work was supported by grants from MOST (2010CB923000, 2011CBA001000, 2011CBA00102, 2012CB821403 and 2013CB921700) and NSFC (11004232, 11034011/A0402, 11234014, 11274362 and 11474330) of China.

%\bibliography{biblio_short}
\bibliography{biblio_long}

\end{document}